\documentclass[12pt,a4paper]{article}

\usepackage[margin=1in]{geometry}
\usepackage{setspace}
\usepackage{amsmath,amssymb,amsfonts,bm}
\usepackage{graphicx}
\usepackage{xcolor}
\usepackage{hyperref}
\usepackage[numbers,sort&compress,square]{natbib}

\hypersetup{
    colorlinks=true,
    linkcolor=blue,
    citecolor=blue,
    urlcolor=blue
}

\begin{document}


\begin{center}

{\Large \textbf{Achieving ultrahigh-temperature superfluidity in atomic Fermi gases by suppressing pairing fluctuations}}\\[1.5em]


Lin Sun,\textsuperscript{1} and
Qijin Chen,\textsuperscript{1,2,3}*\\[1em]


\textsuperscript{1}Hefei National Laboratory, Hefei 230088, China\\
\textsuperscript{2}Hefei National Research Center for Physical Sciences at the Microscale and School of Physical Sciences, University of Science and Technology of China, Hefei 230026, China\\
\textsuperscript{3}Shanghai Research Center for Quantum Science and CAS Center for Excellence in Quantum Information and Quantum Physics, University of Science and Technology of China, Shanghai 201315, China\\[0.5em]



*Correspondence: qjc@ustc.edu.cn

\end{center}

\vspace{1em}


\section*{PUBLIC SUMMARY}

\begin{itemize}
    \item Strong attraction can bind fermions into pairs, but pairing
    fluctuations often prevent superfluidity from surviving at high
    temperatures.

    \item We predict that combining a three-dimensional optical lattice
    with a small imbalance between the two spin populations strongly
    suppresses these fluctuations.

    \item Momentum confinement enhances Pauli blocking between paired
    and excess fermions, raises pair-excitation energies, reduces the
    pseudogap, and dramatically increases the superfluid transition
    temperature.

    \item Under the conditions studied, the enhancement is optimized at
    an imbalance of about 10\% and is accessible to current ultracold-atom
    experiments.

    \item A gapped pair spectrum and unusually weak thermal depletion of
    the condensate provide measurable signatures, while the underlying
    mechanism may guide searches for higher-temperature superconductivity
    in solids.
\end{itemize}


\begin{abstract}
       Achieving room-temperature superconductivity has been a
       long-sought goal of both the physics community and the
       industry. The superfluid transition temperature $T_\mathrm{c}$
       of a Fermi system initially increases with increasing pairing
       strength between fermions following the mean-field prediction
       of the Bardeen-Cooper-Schrieffer (BCS) theory, and is then
       suppressed by pairing fluctuations as the strength increases
       further, which prevents $T_\mathrm{c}$ from becoming high. Here
       we propose a mechanism for suppressing pairing fluctuations
       using the synergy between population imbalance and momentum
       space confinement in lattice systems, enabling an ultrahigh
       $T_\mathrm{c}$ at moderately strong attraction.  We show that
       in a three-dimensional optical lattice, strong Pauli exclusion
       between paired and excess majority fermions pushes up the pair
       excitation energy and suppresses finite-momentum pairing
       fluctuations. The mechanism is general to any lattice with a
       finite Brillouin zone; the optimal imbalance is modest ($p\lesssim 
       0.1$), keeping the Zeeman field small and avoiding orbital
       complications. The gapped pair dispersion at moderate imbalance
       exponentially suppresses thermal condensate depletion,
       providing a testable signature. The gapped dispersion, enhanced
       condensate fraction, and pseudogap reduction are all accessible
       in present-day experiments. This work opens new avenues toward
       room-temperature superconductivity.
\end{abstract}


\section{INTRODUCTION}

Due to its huge potential for industrial applications,
room-temperature superconductivity at ambient pressure has been a
dream for both the academia and the industry.  This pursuit
intensified after the discovery of the cuprate superconductivity in
1986, which dramatically raised the superconducting transition
temperature $T_\mathrm{c}$ from typically a few kelvins up to 95~K for
optimally doped YBa$_2$Cu$_3$O$_{7-\delta}$ and
Bi$_{2}$Sr$_{2}$CaCu$_{2}$O$_{8+\delta}$ at ambient pressure
\cite{Timusk1999}.  A great many efforts have been put into the
competition of raising $T_\mathrm{c}$ in existing or new families of
superconductors.  For instance, $T_\mathrm{c}$ was further raised up
to 164~K for the Hg-based cuprates under high pressure \cite{Chu1994}.
More recently, under an enormous high pressure of 170 GPa, a record
high $T_\mathrm{c}=250$~K was reported in LaH$_{10}$
\cite{Drozdov2019}.  The intense competition has even stimulated a few
uncorroborated or false claims of discoveries of near room-temperature
superconductivity at high \cite{Snider2020,Dasenbrock-Gammon2023} or
near ambient pressure \cite{Garisto2023a}.

Although many new superconductors were discovered without prior
theoretical predictions, it would be ideal if one can design a way to
realize ultrahigh transition temperatures.  Typically, the transition
temperature as predicted by the BCS mean-field theory may increase
indefinitely with an increasing pairing interaction strength.  It is
now known such an increase is unfortunately suppressed by the pairing
fluctuations that emerge as the interaction becomes strong.
Therefore, it would raise $T_\mathrm{c}$ if one could suppress such
fluctuations.

Here we propose a new mechanism that uses the synergy between a
population imbalance and suitable lattice confinement of the charge
carriers in the momentum space to effectively suppress pairing
fluctuations.  We note that either population imbalance or lattice
confinement alone is normally detrimental to pairing and hence
superfluidity, yet when combined, the two together can lead to a
dramatic increase in $T_\mathrm{c}$.  Such a mechanism can be
illustrated and tested in ultracold atomic Fermi gases in a
three-dimensional (3D) optical lattice with an on-site attractive
interaction and a population imbalance.

In the context of superfluidity, ultracold Fermi gases can be regarded
as the charge-neutral analog of superconductors, which have provided
an ideal platform for investigating the pairing and superfluid physics
over the past decades, primarily owing to their high tunability of
multiple parameters \cite{chen2005PR,bloch2008RMP}.  Using a Feshbach
resonance \cite{chin2010RMP}, one can continuously tune the effective
pairing strength from the weak coupling BCS limit to the strong
pairing Bose-Einstein condensation (BEC) limit.  There have been a
great number of experimental and theoretical studies on ultracold
Fermi gases in recent years.  Many tunable parameters have been made
accessible experimentally, including pairing interaction strength
\cite{chen2005PR,zwerger2011}, population imbalance
\cite{zwierlein2006S,partridge2006S,chien2006PRL,radzihovsky2010RoPP},
optical lattices \cite{bloch2005NP,kohl2005PRL,schneider2008S},
dimensionality \cite{martiyanov2010PRL,wu2015PRL}, etc.  The
suppression of $T_\mathrm{c}$ from its mean-field counterpart
$T_\mathrm{c}^\mathrm{MF}$ normally originates from the emergence of a
pairing-fluctuation-driven pseudogap between these two temperatures.
This has provided a compelling explanation for the widespread
pseudogap phenomena in the high-temperature superconductors
\cite{Janko,chen1998PRL,SadeMelo,arpesanl,Yao2024Nature}.

It is widely recognized that a population imbalance suppresses or
destroys superfluidity in 3D homogeneous or trapped Fermi gases
\cite{chien2006PRL,radzihovsky2010RoPP}.  Only intermediate
temperature superfluids exist at unitarity and in the BCS regime, with
superfluidity completely destroyed at zero temperature
\cite{chien2006PRL,HeLY2006PRB}.  While a stable polarized superfluid
exists in the BEC regime with a finite imbalance $p$, population
imbalance serves to suppress $T_\mathrm{c}$.  Meanwhile, the lattice
periodicity is also known to strongly suppress $T_\mathrm{c}$, in
addition to the reduced fermion mobility; for the balanced case in a
3D lattice, $T_\mathrm{c}$ decreases with the on-site attractive
interaction ($U<0$) as $T_\mathrm{c} \propto -t^2/U$ in the BEC
regime, due to the ``virtual ionization'' during the pair hopping
process \cite{NSR,chen1999PRB}.  This makes $T_\mathrm{c}$ very low
and thus the superfluid phase in a lattice Fermi gas hard to access in
the BEC regime.

Although the superfluid transition has not been realized in 3D
lattices so far, superfluidity and pairing phenomena have been
explored experimentally in 2D and 1D optical lattices
\cite{Sommer2012PRL,Liao2010nature,pagano2014NP,revelle2016PRL,Cheng2016PRA}
in the presence of one or two continuum dimensions.  Considering that
either population imbalance or lattice periodicity alone suppresses
$T_\mathrm{c}$, it seems intuitive to assume that superfluidity would
be suppressed in the BEC regime in the presence of both population
imbalance and lattice effects.  However, we show below that in the
presence of momentum confinement in a three-dimensional optical
lattice (3DOL), $T_\mathrm{c}$ can be dramatically \emph{enhanced}
rather than suppressed in the BEC regime by a small population
imbalance.  Unlike the 3D continuum case, the fermion momenta in 3D
lattices are strongly confined within the first Brillouin zone (BZ).
This confinement leads to a significantly large occupation from the
paired fermions for the quantum state of a given momentum across the
entire BZ, and thus a strong Pauli exclusion between paired and excess
majority fermions in the presence of a population imbalance.  In other
words, the excess majority fermions exert a repulsion on
finite-momentum pairs and thus push up the pair excitation energy,
making it more costly to excite such pairs.
In return, this effectively suppresses the pairing fluctuations. 

The physical origin of this effect is the confinement of the fermion momentum distribution
to a finite BZ, which is a general property of any lattice with full 3D periodicity. This
confinement leads to the Pauli blocking mechanism described below.

In this paper, we propose the synergy between population imbalance and
momentum confinement in a 3DOL as a new mechanism for suppressing
pairing fluctuations, and hence to achieve ultrahigh temperature
superfluid transition.  With a proper choice of parameters, we find
that in a population imbalanced Fermi gases in a 3DOL, $T_\mathrm{c}$
may dramatically increase in the BEC regime.  This is caused by the
Pauli blocking effect in the momentum space between paired and excess
majority fermions.


\section{THEORETICAL FORMALISM}

Here we consider a two-component ultracold atomic Fermi gas in a 3D
cubic optical lattice with an attractive on-site pairing interaction,
$U < 0$, as in a simple attractive Fermi Hubbard model.  The
tight-binding band dispersion is given by $\xi_{\mathbf{k}\sigma}
\equiv \epsilon_\mathbf{k}-\mu_\sigma=2t(3-\cos k_x-\cos k_y-\cos
k_z)-\mu_\sigma$, where $t$ is the hopping integral, and $\mu_\sigma$
is the chemical potential for (pseudo)spin $\sigma$, with
$\sigma=\uparrow,\downarrow$ denoting the majority and minority
components, respectively, and $\bar{\sigma}$ for the opposite spin.
The quasi-momentum $k_{i=x,y,z}$ is restricted in the first BZ, and
the lattice constant is set to unity so that $k_i \in (-\pi,\pi]$.  In
  natural units ($\hbar = k_\mathrm{B} = 1$), this framework captures
  how lattice confinement and population imbalance $p$ interplay to
  influence pairing at finite temperatures, with
  $p=(n_\uparrow-n_\downarrow)/(n_\uparrow+n_\downarrow)$ and
  $n_\sigma$ the fermion number density for spin $\sigma$.

We use a pairing fluctuation theory, previously developed for the
pseudogap physics in the cuprates \cite{chen1998PRL} and extended to
the BCS-BEC crossover in ultracold atomic Fermi gases
\cite{chen2005PR}.  The theory goes beyond the BCS mean-field
approximation by self-consistently including finite-momentum pairing
correlations in the single-particle self energy $\Sigma_\sigma(K) =
\Sigma_{\mathrm{sc},\sigma}(K) + \Sigma_{\mathrm{pg},\sigma}(K)$. The
fermions acquire self energy via pair binding and unbinding
processes. Unlike in the BCS mean-field theory, finite-momentum pairs
make a substantial contribution when the pairing interaction is
strong, which leads naturally to a pseudogap in the single-particle
excitation spectrum \cite{chen2005PR,Timusk1999,Yao2024Nature}.  While
the standard Cooper pair condensate contribution,
$\Sigma_{\mathrm{sc},\sigma}(K) = -\Delta_\mathrm{sc}^2
G_{0\bar{\sigma}}(-K)$, vanishes above $T_\mathrm{c}$, the
finite-momentum pair contribution, $\Sigma_{\mathrm{pg},\sigma}(K) =
\sum_{Q\ne 0} t_\mathrm{pg}(Q)G_{0\bar{\sigma}}(Q-K)$, persists both
above and below $T_\mathrm{c}$.
Here $\Delta_\mathrm{sc}$ is the superconducting order parameter, 
$G_{0\sigma}(K)=(\mathrm{i}\omega_{n}-\xi_{\mathbf{k}\sigma})^{-1}$ is
the bare Green's function, and $t_\mathrm{pg}(Q)$ is the pairing
$T$-matrix.  A comparison with alternative $T$-matrix approaches
\cite{Tamaki2008,Frank2018,Kawamura2022} is provided in the
Supplemental Materials.  The four-momentum notation for $K$ and $Q$, as well as
further details about this theory, is described in the Supplemental
Materials.

The superfluid transition temperature is determined by the Thouless
criterion, where the inverse pair propagator $t_\mathrm{pg}^{-1}(0) =
0$, which is equivalent to the pair chemical potential $\mu_\mathrm{p}
= 0$ at $T \leq T_\mathrm{c}$.  This leads to the pseudogap
approximation, $\Sigma_{\mathrm{pg},\sigma}(K) \approx
-\Delta_\mathrm{pg}^2G_{0\bar{\sigma}}(-K)$, which defines the
pseudogap parameter $\Delta_\mathrm{pg}$ via
$\Delta_\mathrm{pg}^2\approx -\sum_{Q\ne 0} t_\mathrm{pg}(Q)$. Then we
arrive at the self-energy,
$\Sigma_\sigma(K)=-\Delta^2G_{0\bar{\sigma}}(-K)$, and the full
Green's function,
$G_\sigma(K)=u_{\mathbf{k}}^{2}/(\mathrm{i}\omega_{n}-E_{\mathbf{k}\sigma})+v_{\mathbf{k}}^{2}/(\mathrm{i}\omega_{n}+E_{\mathbf{k}\bar{\sigma}})$,
both in simple BCS-like forms, with the total gap
$\Delta=\sqrt{\Delta^{2}_\mathrm{sc}+\Delta^{2}_\mathrm{pg}}$, and
coherence factors $u_{\mathbf{k}}^2 = (1 +
\xi_{\mathbf{k}}/E_{\mathbf{k}})/2$, $v_{\mathbf{k}}^2 = (1 -
\xi_{\mathbf{k}}/E_{\mathbf{k}})/2$, where $E_{\mathbf{k}} =
\sqrt{\xi_{\mathbf{k}}^2 + \Delta^2}$.  The quasiparticle energies are
$E_{\mathbf{k}\uparrow} = E_{\mathbf{k}} - h$ and
$E_{\mathbf{k}\downarrow} = E_{\mathbf{k}} + h$, with $\mu =
(\mu_\uparrow + \mu_\downarrow)/2$ and $h = (\mu_\uparrow -
\mu_\downarrow)/2$. Naturally, this also leads to a BCS-like gap
equation (See Supplemental Materials for more details), except that a
pseudogap is already present at $T_\mathrm{c}$, which necessarily
suppresses $T_\mathrm{c}$ from its mean-field counterpart.

\begin{figure}[h]
	\centering
	\includegraphics[width=0.6\textwidth,clip]{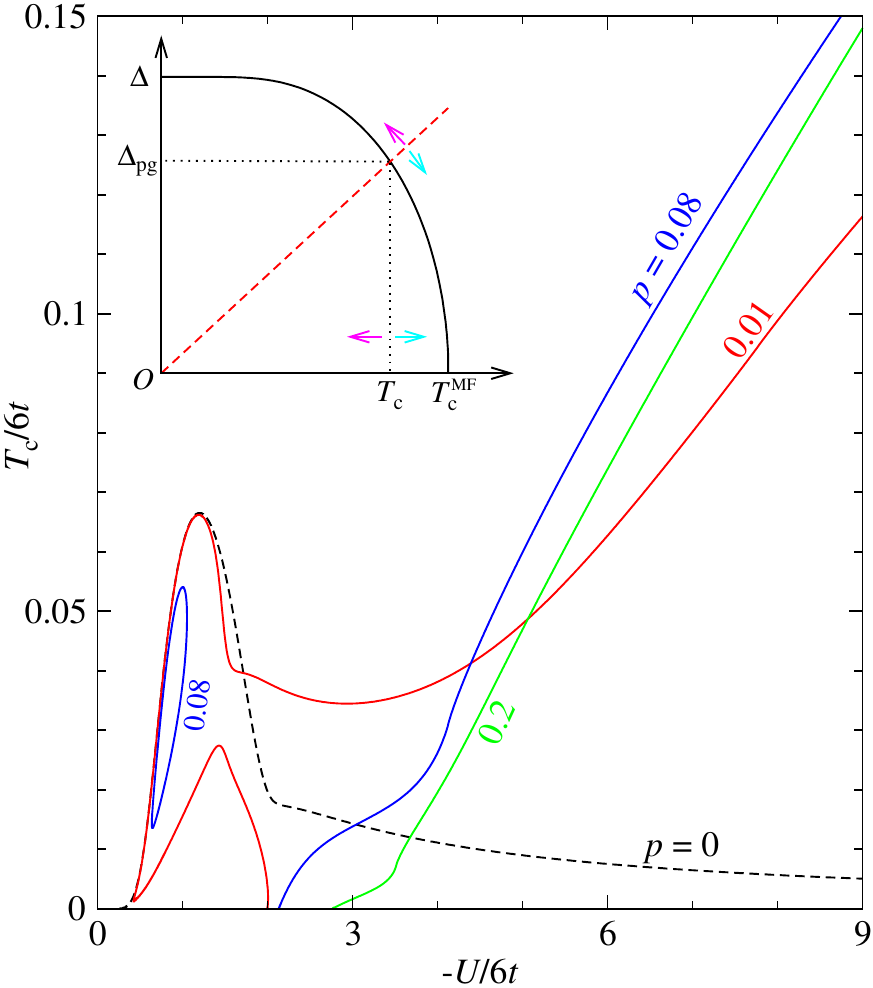}
	\caption{\textbf{Population imbalance enhances the superfluid transition temperature in the strong-coupling regime.}
          Evolution of $T_\mathrm{c}$ as a function of
          interaction strength $-U/6t$ for $n=0.15$ with varying $p$,
          as labeled.  Also plotted for comparison is the balanced
          case (black dashed line).  $T_\mathrm{c}$ is dramatically
          raised in the BEC regime by population imbalance. Shown
          schematically in the inset is how increasing the pseudogap
          $\Delta_\mathrm{pg}$ suppresses $T_\mathrm{c}$ or vice
          versa. The total gap $\Delta(T)$ (black solid line) roughly
          follows the BCS mean-field behavior, while $\Delta
          _\mathrm{pg}$ increases with $T$ (red dashed line).}
	\label{fig:Fig1}
\end{figure}

The inset of figure \ref{fig:Fig1} illustrate schematically how
pairing fluctuations, and hence the pseudogap, suppress
$T_\mathrm{c}$. The overall gap $\Delta$ roughly follows the BCS
mean-field relation as a function of temperature. $T_\mathrm{c}$ is
determined by the intersecting point between the curves of $\Delta(T)$
(black solid line) and $\Delta_\mathrm{pg}(T)$ (red dashed
line). Therefore, as the pseudogap increases, this point will move
along the $\Delta(T)$ curve toward lower temperature, leading to
suppression of $T_\mathrm{c}$. On the contrary, when pairing
fluctuations are suppressed (via the mechanism we propose here),
$\Delta_\mathrm{pg}$ decreases, and $T_\mathrm{c}$ will rise toward
its mean-field value, $T_\mathrm{c}^\mathrm{MF}$.

The size of the pseudogap is determined by integrating the occupied
spectrum of finite-momentum pairs. The pair dispersion is given by the
poles of the pair propagator, and can be approximated as
$\tilde{\Omega}_{\mathbf{q}}\approx \Omega_{\mathbf{q}} =
2t_\mathrm{B} (3 - \cos q_x - \cos q_y - \cos q_z)$ in the balanced
case at $T_\mathrm{c}$, where $t_\mathrm{B}$ represents the effective
pair hopping matrix element.  It uniquely determines the pseudogap at
$T_\mathrm{c}$, $\Delta_\mathrm{pg}^2 \simeq
\frac{1}{2|a_0|}\sum_\mathbf{q}
\left(\coth\frac{\Omega_{\mathbf{q}}}{2T} -1\right)$, where $a_0$ is
the inverse residue of the pair propagator (See the Supplemental
Materials for further information).  Using this relation, along with
the number constraint, $n_\sigma = \sum_K G_\sigma(K)$, we solve the
$T_\mathrm{c}$ equation self-consistently for $\mu_\sigma$,
$\Delta_\mathrm{pg}$, and $T_\mathrm{c}$ (with $\Delta_\mathrm{sc} =
0$).  The parameters where the $T_\mathrm{c}$ enhancement is largest
correspond to the shallow BEC regime, where $|\mu|$ remains comparable
to the bandwidth and a perturbative expansion in $\Delta/|\mu|$ is not
appropriate.

It should be emphasized that the bare Green's function $G_0$ is not
completely ``bare'', as the chemical potential always contains
interaction effects. In the $G_0G$ approach derived from equations of
motion \cite{Kadanoff1961,ChenPhD}, the Hartree self-energy is
absorbed into the chemical potential. More generally, $G_0$ is
constructed to contain the diagonal self-energy contributions such
that $G_0^{-1}(\omega=0,\mathbf{k}_\mu)=0$, where $\mathbf{k}_\mu$ is
the anisotropic wavevector on the Fermi surface. In other words, the
difference between $G(K)$ and $G_0(K)$ is only the off-diagonal
self-energy. Our pseudogap approximation for the self-energy keeps
precisely this off-diagonal part, with the diagonal part absorbed
inside $\mu_\sigma$ (which differs for the two spins in the presence
of imbalance). Consequently, the pseudogap approximation automatically
satisfies the above requirement. Note that $\mu_\sigma$ differs from
the physical chemical potential by the absorbed diagonal self-energy,
even though it yields the correct Fermi wavevector. In this way, our
$G_0G$ formalism is free of the artificial imbalance that some may
think \cite{Pini2019}.


\section{NUMERICAL RESULTS AND DISCUSSIONS}

\subsection{Behavior of $T_\mathrm{c}$ versus $U$ for $p\ne 0$}

We first present in figure \ref{fig:Fig1} the behaviors of
$T_\mathrm{c}$ as a function of $-U/6t$ at a representative density
$n=0.15$ for a series of values of population imbalance $p$, as
labeled.  Here $6t$ is the half bandwidth.  The threshold coupling for
forming a zero-binding-energy two-body bound state is given by
$U_c/6t\approx -1.319$, which is the counterpart of the unitarity in
3D free space. As a reference for comparison, we also plot the
balanced $p=0$ curve (black dashed), which sees an expected decreasing
$T_\mathrm{c}$ in the BEC regime, following the scaling $T_c\propto
-t^2/U$.

On the BCS side of unitarity, there exists a superfluid phase at
intermediate temperatures for a narrow range of small $p\ne 0$, due to
the Fermi surface mismatch between the majority and minority fermions
\cite{chien2006PRL}.  The intermediate temperature superfluid phase in
the BCS regime shrinks and eventually disappears as $p$ increases
beyond 0.08. Indeed, as one can intuitively expect, a sufficiently
large Fermi surface mismatch will render zero-momentum pairing
impossible at relatively weak interactions.

On the BEC side, the $T_\mathrm{c}$ curve for $p=0.01$ sees a much
smaller drop compared with the $p = 0$ case, indicating that a nonzero
$p$ will substantially raise $T_\mathrm{c}$ in the BEC regime, which
is also reported in a 1D optical lattice system
\cite{wang2020PRA2}. For higher $p$, the $T_\mathrm{c}$ curves also
revive in the BEC regime, where the Fermi surface no longer exists,
and the fermion momentum spreads across the entire BZ.  Remarkably,
for all the $p\ne 0$ cases shown, $T_\mathrm{c}$ rapidly increases as
the interaction strength goes stronger beyond $-U/6t\approx 3$.  An
ultrahigh superfluid transition temperature of atomic Fermi gases in
the 3DOL can be achieved in the BEC region with population imbalance.

\subsection{Suppression of Pairing Fluctuations}

Such extraordinary behaviors of $T_\mathrm{c}$ in the BEC regime can
be easily understood via the momentum distributions of the paired
fermions and the excess majority fermions $\delta
n_{\mathbf{k}}=n_{\mathbf{k}\uparrow}-n_{\mathbf{k}\downarrow}$.  For
$p\ne 0$ in the BEC regime, the momentum distribution of the minority
fermions $n_{{\mathbf{k}}\downarrow}$ roughly reflects that of paired
fermions, (which would be given by the coherence factor
$v_\mathbf{k}^2$, had there been no imbalance).  As is well known, in
the 3D continuum case in the deep BEC regime, the momentum
distribution of ``paired fermions'' $v_\mathbf{k}^2$ will spread to
the entire infinitely large momentum space and goes to nearly zero,
and thus the excess majority fermions feel no Pauli blocking from
paired fermions or vice versa.  In contrast, in the present 3DOL case
with a population imbalance, the distributions of both the excess
majority fermions and the paired fermions still remain finite where
they overlap and do not get small in the BEC limit, as they are
restricted to the first BZ. This leads to a strong repulsion between
them due to Pauli blocking.  This repulsion increases with population
imbalance, interaction strength and density, and decreases with
temperature.  The repulsion from excess majority fermions pushes up
the energy of fermion pairs, and thus making them harder to excite. As
a result, pairing fluctuations are effectively suppressed and the
pseudogap is reduced, leading to a dramatic rise of $T_\mathrm{c}$.

\begin{figure}
	\centering
	\includegraphics[width=0.9\textwidth,clip]{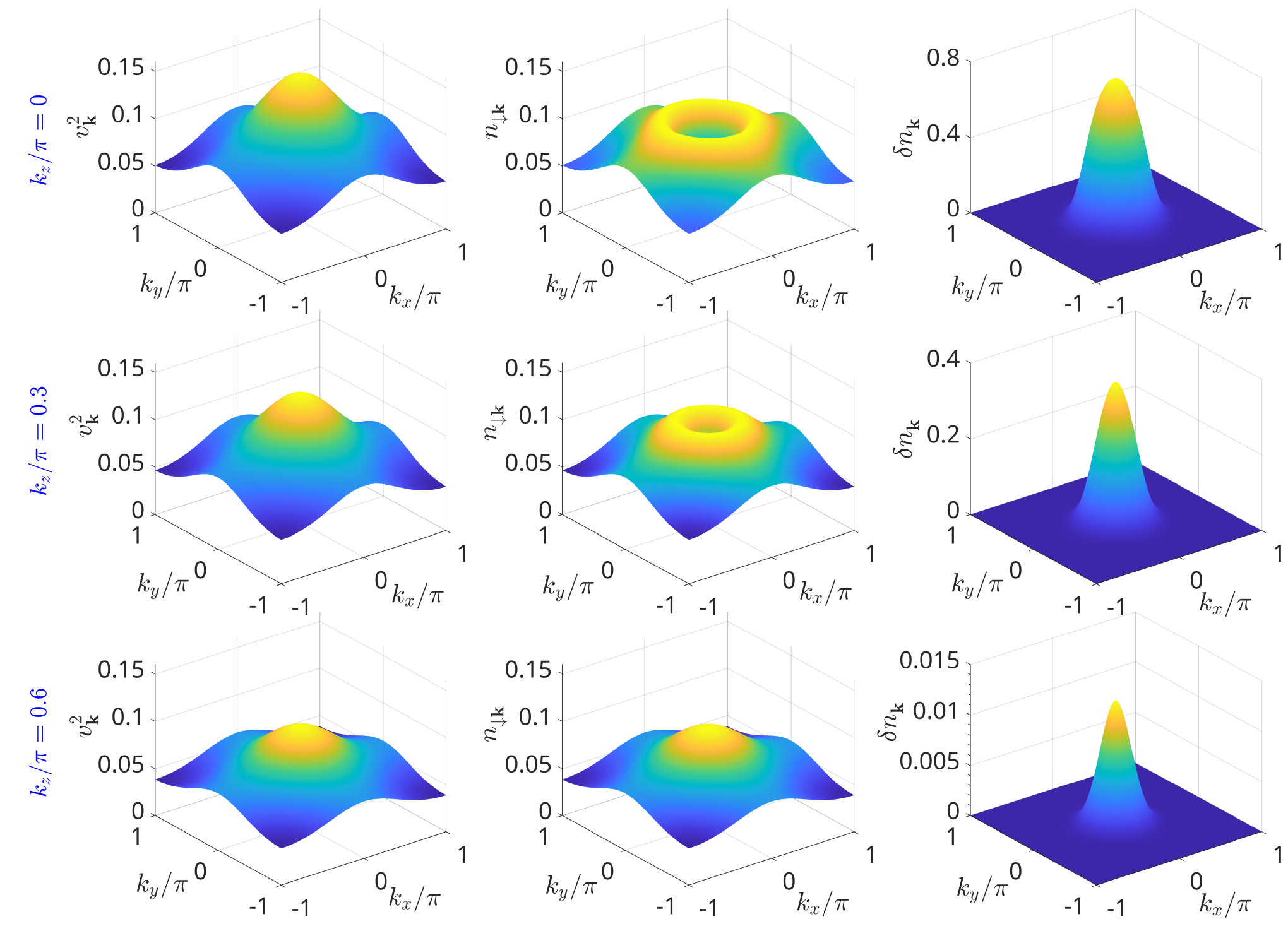}
	\caption{\textbf{Pauli exclusion redistributes paired and excess fermions in momentum space.}
          Momentum distributions of $v_\mathbf{k}^2$ (left),
          $n_{\mathbf{k}\downarrow}$ (middle) and $\delta
          n_\mathbf{k}$ (right column) at $T_\mathrm{c}$ in the
          $(k_x,k_y)$ plane with $k_z/\pi = 0$ (top), 0.3 (middle) and
          0.6 (bottom row), for $-U/6t=5$, $p=0.1$, and $n=0.15$.  A
          hole is dug out at the center for
          $n_{\mathbf{k}\downarrow}$, due to the Pauli repulsion with
          the excess majority fermions $\delta n_\mathbf{k}$, which is
          peaked at $\mathbf{k}=0$.  }
	\label{fig:Fig2}
\end{figure}

To understand how the repulsion between excess majority and paired
fermions occurs, we investigate the momentum distributions of the
excess majority and the minority fermions (hence the paired fermions)
in the BEC regime.  Presented in figure \ref{fig:Fig2} are
representative 3D surface plots of $v_\mathbf{k}^2$ (left),
$n_{\mathbf{k}\downarrow}$ (middle) and $\delta n_\mathbf{k}$ (right
column) at $T_\mathrm{c}$ in the $(k_x,k_y)$ plane at $k_z/\pi = 0$
(top), $0.3$ (middle) and $0.6$ (bottom row), respectively, for
$n=0.15$ with $p=0.1$ at $-U/6t=5$, corresponding to a high
$T_\mathrm{c}/6t \approx 0.06$ in figure \ref{fig:Fig1}. While the
variation in the $(k_x, k_y)$ plane is directly displayed in each
plot, the change in the $\hat{k}_z$ direction can be seen from the top
to the bottom row. For example, $\delta n_\mathbf{k}$ decreases with
$|k_z|$, so that $n_{\mathbf{k}\downarrow}\approx v_\mathbf{k}^2$ at
large $|k_z|$.

The excess majority fermions $\delta n_{\mathbf{k}}$ (right column) is
peaked at zero momentum, in all momentum directions. At the same time,
the momentum distribution of paired fermions is dug out at the low
momentum center, leaving a hole in $n_{\downarrow{\mathbf{k}}}$
(middle column), as compared to $v_\mathbf{k}^2$ (left column). This
is a result of Pauli exclusion between paired and excess majority
fermions. It is this mutual exclusion that requires that, when pairs
form, the pre-occupied low momentum states for the constituent
fermions now have to be moved to higher energy. Therefore, the fermion
pairs will acquire substantially higher energy at give $\mathbf{k}$,
as compared to its balanced counterpart, making them harder to excite.

\begin{figure}
	\centering
	\includegraphics[width=0.6\textwidth,clip]{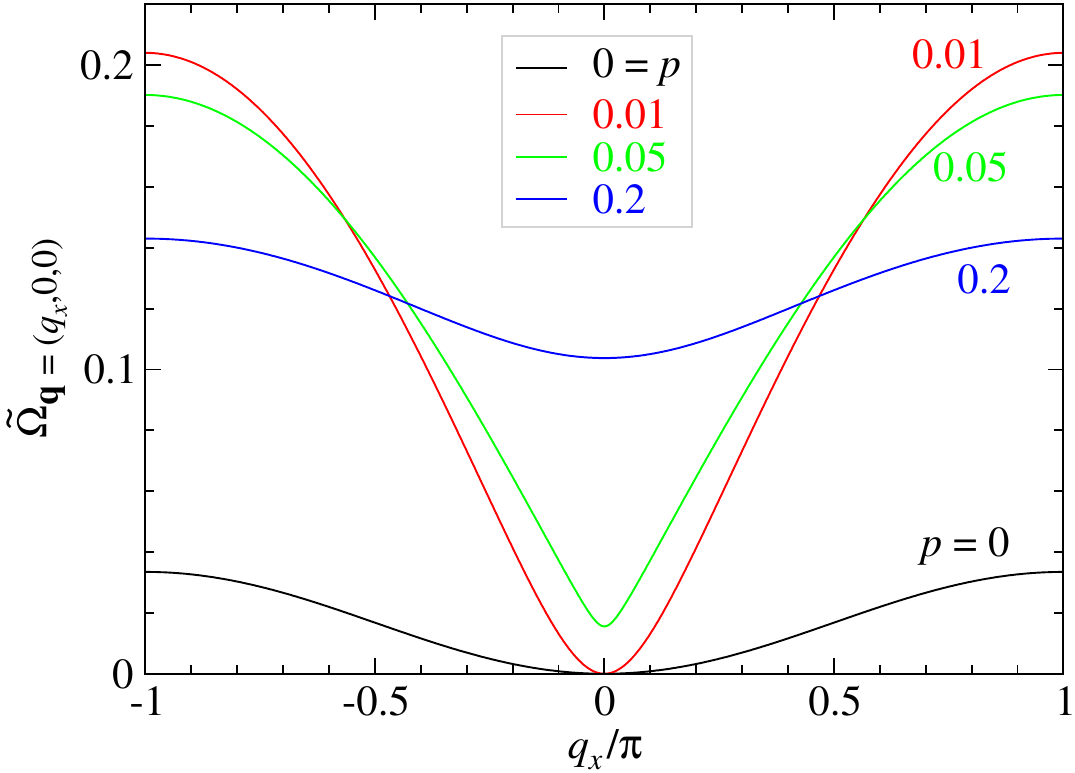}
	\caption{\textbf{Population imbalance raises the pair-excitation energy and opens a dispersion gap.} 
          Pair dispersion $\tilde{\Omega}_{\mathbf{q}=(q_x,0,0)}$ at $T_\mathrm{c}$
          along the $q_x$ direction for different $p$, as labeled, for
          $n=0.15$ and $-U/6t=5$ (in the BEC regime). An excitation
          gap emerges in $\tilde{\Omega}_{\mathbf{q}}$ as $p$
          increases beyond 0.05. The pair energy is dramatically
          elevated by population imbalance, as compared to the $p=0$
          case, leading to strong suppression of pairing
          fluctuations. }
	\label{fig:Fig3}
\end{figure}

The effect of Pauli exclusion can be seen in figure \ref{fig:Fig3},
which presents the pair dispersion
$\tilde{\Omega}_{\mathbf{q}=(q_x,0,0)}$ with momentum along the $q_x$
direction only for different $p$'s at $T_\mathrm{c}$ for $n=0.15$ and
$-U/6t=5$.  Compared to the $p=0$ case (black curve), the pair energy
for $p\ne 0$ is dramatically elevated. For $p=0.05$ and above, an
excitation gap at $\mathbf{q}=0$ emerges in pair dispersion
$\tilde{\Omega}_{\mathbf{q}}$; the overall pairing becomes hole-like,
with $a_0 < 0$.  We find that low temperature, strong interaction and
large population imbalance, together with high density, add to the
increase of $\delta n_{\mathbf{k}}$ at low momenta, causing stronger
Pauli repulsion and hence suppression of pairing fluctuations in the
BEC regime. As a result, the pseudogap is reduced and $T_\mathrm{c}$
is dramatically raised in the BEC regime toward its mean-field value
$T_\mathrm{c}^\mathrm{MF}$.

\begin{figure}[h]
	\centering
	\includegraphics[width=0.95\textwidth,clip]{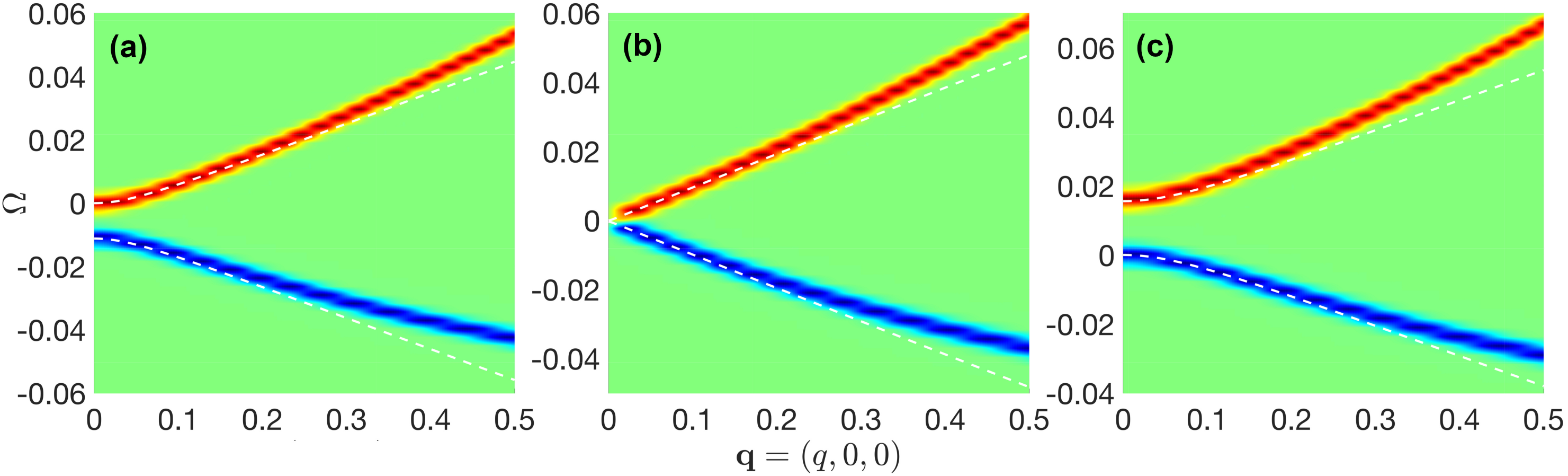}        
	\caption{\textbf{Pair spectra evolve from gapless to gapped with increasing population imbalance.}
          Spectral function of fermion pairs,
          $B(\mathbf{q},\Omega)$, along the $\Gamma-X$ direction for
          $n=0.15$ and (a) $p=0.03$, (b) 0.037, and (c) 0.05,
          corresponding to $a_0 >0$, $a_0 = 0$ and $a_0 < 0$,
          respectively. The white dashed lines are the pair
          dispersions obtained from the Taylor expansion of
          $t_{\textrm{pg}}^{-1}(\mathbf{q},\Omega)$, with the positive
          branch giving $\tilde{\Omega}_\mathbf{q}$.}
	\label{fig:Fig4}
\end{figure}

It should be noted that, when $a_0<0$, the Thouless criterion is still
satisfied, even though the positive frequency branch of the pair
dispersion appears gapped. As shown in Fig.~\ref{fig:Fig4} for
$-U/6t=5$ and $n=0.15$, the pair spectral function
$B(\mathbf{q},\Omega)\equiv -2
\textrm{Im}\,t_\textrm{pg}(\mathbf{q},\Omega)$ reveals that the pair
dispersion contains both a positive and a negative frequency branch,
similar to the two-branch dispersion $\omega = ck$ of an
electromagnetic field or a phonon. Here the positive branch is
associated with the energy for exciting a pair out of the condensate,
whereas the negative branch is the energy gain for a finite-momentum
pair to anihilate into the condensate. In the presence of excess
fermions, excited pairs will feel a Pauli repulsion, leading to an
elevated excitation energy cost. Indeed, both the positive and
negative branches exhibit a gradual up-shift at finite momenta. For
small imbalance, $p=0.03$ ($a_0 > 0$), there is a finite energy gain
even at $q=0$ for the negative branch due to the Pauli repulsion. For
$p=0.037$ ($a_0 = 0$), the dispersions of both branch becomes
photon-like or linear in the long wavelength limit. For higher
imbalance, $p=0.05$ ($a_0 < 0$), the Pauli repulsion is so strong that
it takes a nonzero excitation energy to bring a pair out of the
condensate to an infinitesimally small momentum for the positive
branch. This finite excitation gap does not necessarily destroy the
condensation, similar to the case of a charged Bose gas, which still
undergoes BEC at low $T$ \cite{Foldy1961}. Indeed, the repulsive
interaction between pairs and excess fermions play in some sense the
role of the Coulomb repulsion in the charged Bose gas.

The gapped pair dispersion in Fig.~\ref{fig:Fig4}(c) further enhances
the condensate fraction. In standard BEC theory with a gapless
spectrum $\Omega_{\mathbf{q}} \propto q^2$ (Fig.~\ref{fig:Fig4}(a)),
the thermal depletion follows $n_{\mathrm{pair}}^\textrm{ex} \propto
T^{3/2}$ in 3D. When an excitation gap $\Omega_0 > 0$ opens in
$\tilde{\Omega}_{\mathbf{q}}$, the occupation of finite-momentum pairs
becomes exponentially suppressed,
$n_{\mathrm{pair}}^\textrm{ex} 
\sim e^{-\Omega_0/T}$, at low $T \ll \Omega_0$. Consequently, the
condensate fraction $n_0/n = 1 - 2n_{\mathrm{pair}}^\textrm{ex}/n$
remains near unity up to a much larger fraction of $T_\mathrm{c}$ when
$a_0<0$ than when $a_0>0$. This is the same physical origin as the
$T_\mathrm{c}$ enhancement, but provides a separate experimental
signature: a gapless pair dispersion (when $a_0>0$) gives the familiar
power-law thermal depletion, while a gapped dispersion ($a_0<0$) gives
exponentially suppressed depletion, with the condensate persisting
anomalously close to $T_\mathrm{c}$.

\subsection{Achieving Ultrahigh $T_\mathrm{c}$ with an Optimal Imbalance}

\begin{figure}
	\centering
	\includegraphics[width=0.6\textwidth,clip]{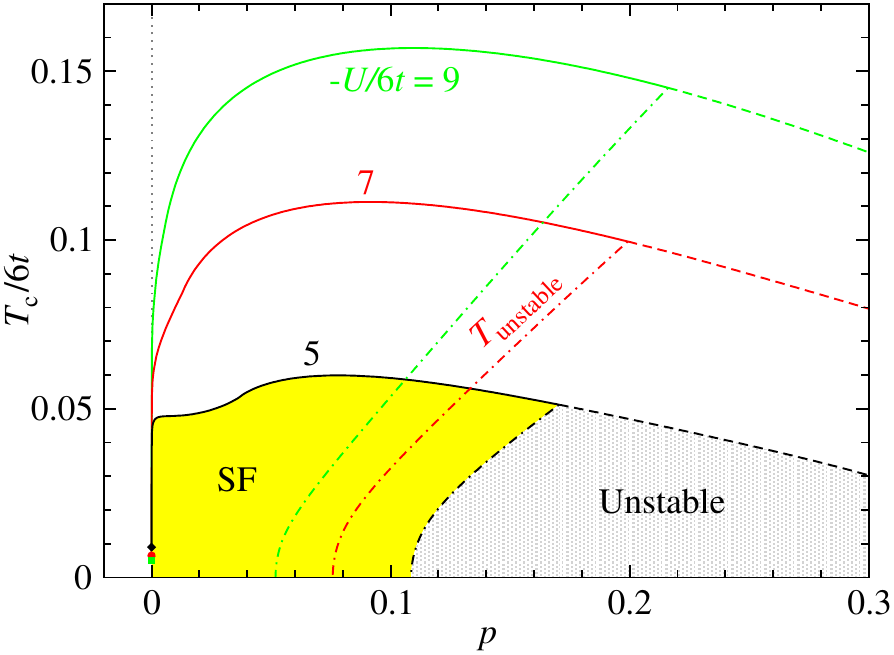}
	\caption{\textbf{A modest population imbalance optimizes the superfluid transition temperature.} 
          Behavior of $T_\mathrm{c}$ versus $p$ for $n=0.15$ at
          $-U/6t=5,7,9$ (as labeled) in the BEC regime.  The
          color-coded symbols near the origin denote the corresponding
          $T_\mathrm{c}$ values for $p=0$.  Here $T_\mathrm{c}$
          undergoes an initial sharp jump from $p\equiv 0$ to $p=0^+$,
          and then reaches a maximum around $p=0.1$, followed by a
          slow decrease as $p$ increases further. The dashed part of
          the $T_c$ curves is unstable, where $p$ is large. The
          dot-dashed lines are instability boundaries given by
          $T_\mathrm{unstable}$, which separate the stable polarized
          superfluid and the unstable phases, shown for $-U/6t=5$ as
          yellow shaded (denoted by ``SF'') and gray dot-patterned
          areas, respectively. }
	\label{fig:Fig5}
\end{figure}

Not all values of imbalance and interaction strength lead to enhanced
$T_\mathrm{c}$.  There is an optimal imbalance $p$ for achieving the
highest $T_\mathrm{c}$, as an overly large $p$ will necessarily
destroy pairing and superfluidity.  This can be seen in
Fig.~\ref{fig:Fig5} from the plot of $T_\mathrm{c}$ as a function of
$p$ for $n=0.15$ in the BEC regime, at $-U/6t = 5, 7, 9$.  There is an
initial sharp jump in $T_\mathrm{c}$ at $p=0^+$, in the very vicinity
of the identically $p=0$ point (denoted by the color-coded symbols).
This discontinuity arises because Pauli exclusion is a binary
constraint, not a finite-strength interaction.  At exactly $p=0$ there
are no excess fermions and the effect is entirely absent; for any
$p>0$ excess majority fermions occupy the zone center and Pauli
repulsion is fully active.  Even an infinitesimal imbalance
immediately displaces paired fermions from low-momentum states,
producing the sharp onset.  For given $U/6t$, $T_\mathrm{c}$ reaches a
maximum around $p=0.1$, followed by a slow decrease as $p$ increases
further, which has largely to do with the decreasing pair density.  In
particular, for $p=0.11$, $T_\mathrm{c}/6t$ rises dramatically from
$0.058$ to an ultrahigh value of $0.157$ as the interaction strength
$-U/6t$ varies from $5$ to $9$. As expected, for large $p$, the
polarized superfluid becomes unstable (indicated by the dashed lines).
The dot-dashed $T_\mathrm{unstable}$ lines, which are calculated
following Ref.~\cite{Chen2006Instability}, separate the stable from
the unstable polarized superfluid phases. In particular, for
$-U/6t=5$, these two phases are indicated as the yellow-shaded
(denoted by ``SF'') and grey dot-patterned areas, respectively.  Phase
separation is likely to occur in the unstable phases.

A noteworthy feature is that the optimal imbalance is modest, $p
\lesssim 0.1$, corresponding to a small Zeeman field $h$.  This
allows the mechanism to be translated to solid-state systems without
the orbital complications (vortices, Landau level quantization) that
arise at high fields.


\section{EXPERIMENTAL PROPOSALS}

The enhanced $T_\mathrm{c}$ and the underlying suppression of pairing
fluctuations can be tested with \textit{in situ} probes that do not
require releasing the lattice potential.

Momentum-resolved rf spectroscopy probes the single-particle spectral
function and can distinguish the condensate gap $\Delta_\mathrm{sc}$
from the pseudogap $\Delta_\mathrm{pg}$
\cite{chen2005PR,Yao2024Nature}.  As $p$ increases into the gapped
pair regime, the pseudogap below $T_\mathrm{c}$ should decrease sharply
while the total gap $\Delta$ remains large, signaling the suppression
of pairing fluctuations.  The rf pulse is applied while the lattice is
on, with only individual transferred atoms detected subsequently,
avoiding the problem that on-site pairs dissociate when the lattice is
switched off.  This is the most direct probe of the mechanism.

In the shallow BEC regime (where the $T_\mathrm{c}$ enhancement is
largest, e.g., $-U/6t \sim 5-9$), pairs are well-formed composite
bosons with size smaller than the lattice spacing but not yet in the
deep BEC limit.  \textit{In situ} Bragg spectroscopy (without turning
off the lattice) probes the dynamic structure factor
$S(\mathbf{q},\omega)$.  At low energy, the response is dominated by
center-of-mass excitations of the pair condensate, which behave as
those of a Bose superfluid.  At higher energy, the pair-breaking
continuum sets in at a threshold set by the total gap $\Delta$.  A
comparison of the pair-breaking edge in the Bragg response with the
single-particle gap from rf spectroscopy can, in principle, reveal the
relative weight of $\Delta_\mathrm{sc}$ and $\Delta_\mathrm{pg}$ below
$T_c$, though cleanly separating these components in a 3D lattice
geometry remains challenging.

In the attractive Hubbard model, the doublon (doubly occupied site)
fraction provides an \textit{in situ} measure of the condensation
energy.  At the optimal parameters for $T_\mathrm{c}$ enhancement as
shown in Fig.~\ref{fig:Fig5}, the system is in the shallow BEC regime
where the doublon fraction is not yet saturated.  Upon entering the
superfluid phase, the steep increase in condensation energy in the
case of $a_0 < 0$, arising from the elevated $T_\mathrm{c}$ and
exponentially suppressed thermal depletion of the condensate, should
produce a correspondingly sharp feature in the doublon fraction across
$T_\mathrm{c}$, which is directly accessible via \textit{in situ}
imaging.  In the deeper BEC regime, the doublon fraction would
saturate and the signal would disappear, but near the optimal
parameters this probe is complementary to rf and Bragg spectroscopy.
\textit{In situ} density imaging can also probe the compressibility
$\kappa = \partial n/\partial\mu$, which exhibits a distinct feature
(e.g., a kink or suppression) across $T_\mathrm{c}$ throughout the
BCS-BEC crossover, as demonstrated at unitarity \cite{Ku2012Science},
providing a thermodynamic complement to the spectroscopic probes.

The required experimental ingredients (a 3D optical lattice, Feshbach
resonance, spin-selective loading, and cooling to degeneracy) are
all routine in current cold atom laboratories
\cite{bloch2005NP,bloch2008RMP,zwierlein2006S}.  The $T_\mathrm{c}$
enhancement itself is the key advance: at optimal parameters,
$T_\mathrm{c}$ enters a range achievable with existing technology,
substantially relaxing the cooling challenge that has so far prevented
access to the 3D lattice superfluid.


\section{CONCLUSION}

In summary, we have proposed a scheme to achieve ultrahigh temperature
superfluidity in atomic Fermi gases in 3D optical lattices, by
coordinating a synergy between population imbalance and momentum space
confinement.  Due to this confinement, paired fermions feel a strong
Pauli repulsion from excess majority fermions, and thus acquire a much
elevated excitation energy.  In return, finite-momentum pairing
fluctuations are strongly suppressed, leading to a much enhanced,
ultrahigh superfluid transition temperature in the BEC regime.

We emphasize that the mutual Pauli exclusion between paired and excess
majority fermions due to the momentum space confinement does not rely
on our specific theory; the wide momentum spread across (but limited
to) the entire first BZ largely reflects the Heisenberg uncertainty
principle due to the small pair size in the BEC regime.  Therefore,
our proposal should be equally valid in other competing
pairing-fluctuation-based theories of superfluidity as well
\cite{Levin2010}.

Several features of this mechanism are noteworthy.  First, it is
general: any lattice with full 3D momentum confinement (cubic, FCC,
BCC, or multiband) should exhibit the same enhancement, since the
physics of overlapping momentum distributions in a finite BZ is
universal.  Second, the optimal imbalance is modest ($p \sim
0.1$), corresponding to a small Zeeman field in solid-state
analogues, so orbital effects are negligible and the simple Zeeman
approximation is valid.  Third, the gapped pair dispersion that
emerges at moderate imbalance ($p \gtrsim 0.04$) exponentially
suppresses the thermal depletion of the condensate, contributing to
the $T_\mathrm{c}$ enhancement and providing a distinct experimental
signature. 

The suppression of pairing fluctuations can be probed by multiple
\textit{in situ} techniques.  RF spectroscopy directly distinguishes the
condensate gap from the pseudogap.  Bragg spectroscopy probes the
pair-breaking threshold and the collective response of the pair
condensate.  Near the optimal parameters, the doublon fraction and
compressibility provide thermodynamic signatures of the enhanced
condensation energy.  All of these probes are accessible with existing
cold atom technology; the $T_\mathrm{c}$ enhancement itself
substantially relaxes the cooling requirement that has so far
prevented access to the 3D lattice superfluid.

This proposal may also serve as a prototype for achieving room
temperature superconductivity in solid state lattices.  Observations
of superconductors in the BEC regime have been reported
\cite{PhysRevB.107.104517,Hashimoto2020,Nakagawa2021,Kim2022}.  With
population imbalance tuned via, e.g., an applied magnetic field, one
may realize ultrahigh temperature superconductivity.

\section*{RESOURCE AVAILABILITY}

\subsection*{Materials availability}
Not applicable.

\subsection*{Data and code availability}
The data and code supporting the findings of this study are available from the corresponding author upon reasonable request.

\section*{FUNDING AND ACKNOWLEDGMENTS}

This work was supported by the Innovation Program for Quantum Science and Technology (Grant No. 2021ZD0301904) of China. 

\section*{AUTHOR CONTRIBUTIONS}
Conceptualization and investigation: L.S., Q.C.;
Methodology,software, data curation and formal analysis: L.S.; 
Validation, project administration and funding acquisition: Q.C.; 
Writing -- original draft: L.S.; 
Writing -- review \& editing: Q.C.

\section*{DECLARATION OF INTERESTS}

The authors declare no competing interests.

\section*{SUPPLEMENTAL INFORMATION}

Supplemental Information includes Supplemental Materials and Methods, one figure, and Supplemental References.




\section*{Figure Titles and Legends}

\vspace{1em}

\vspace{1em}

\vspace{1em}

\vspace{1em}

\end{document}


\section*{Supplemental Materials and Methods}

\subsection*{Overview of the pairing fluctuation theory}

In this work, we employ a pairing fluctuation theory to investigate
finite-temperature effects in ultracold atomic Fermi gases, which
generalizes the Bardeen-Cooper-Schrieffer (BCS) theory to strong
interactions by including pairing fluctuation effects
self-consistently. Originally developed to address the pseudogap physics
in cuprate superconductors~\cite{chen1998PRL}, this theory has been
extended to study the BCS-Bose-Einstein condensation (BEC) crossover in
ultracold Fermi gases~\cite{chen2005PR}.

The two-component ultracold Fermi gas in a 3D optical lattice considered
in the present work can be described by an attractive Hubbard model
but with a population imbalance,
\begin{equation}
  H-\sum_\sigma\mu_\sigma N_\sigma=\sum_{\mathbf{k} \sigma} \xi_{\mathbf{k},\sigma} c_{\mathbf{k} \sigma}^{\dagger} c_{\mathbf{k} \sigma}
  + U\sum_{\mathbf{k} \mathbf{k}^{\prime} \mathbf{q}} c_{\mathbf{k}+\mathbf{q} / 2 \uparrow}^{\dagger} c_{-\mathbf{k}+\mathbf{q} / 2 \downarrow}^{\dagger} c_{-\mathbf{k}^{\prime}+\mathbf{q} / 2 \downarrow} c_{\mathbf{k}^{\prime}+\mathbf{q} / 2 \uparrow}\,.
\end{equation}
Here momentum $\mathbf{k} = (k_x, k_y, k_z)$ is restricted to the first BZ, with the
lattice constant set to unity.  Unlike BCS theory, which assumes zero
center-of-mass momentum ($\mathbf{q}=0$) for fermion pairs, the
contribution of nonzero $\mathbf{q}$ pairing is included
self-consistently and becomes progressively important as the pairing
interaction becomes stronger, and naturally leads to the formation of
a pseudogap.

Our pairing fluctuation theory is derived using an equation of motion
approach; equations of motion were derived for both single-particle
Green's function $G(K)$ and two-particle Green's function
$G_2(Q)$. The equation for $G_2$ is then truncated at the (connected)
three-particle level. This necessarily includes the contribution of
pairing fluctuations in the single-particle self energy $\Sigma(K)$,
which in return gives rise to the formation of a pseudogap in the
single-particle energy spectrum when the pairing interaction becomes
strong. As shown in the inset of Fig. 1, $T_\text{c}$ will then be suppressed
from its mean-field value by the pseudogap. Throughout this paper, we
use a four-momentum notation: $K \equiv (\omega_{n}, \mathbf{k})$, $Q
\equiv (\Omega_{n}, \mathbf{q})$, $\sum_Q \equiv
T\sum_n\sum_\mathbf{q}$, with $\omega_n$ ($\Omega_n$) being the odd
(even) Matsubara frequency.

The dispersion of fermion pairs is determined by the poles of $G_2$, or equivalently by the particle-particle scattering $T$ matrix, $t(Q)$, which contributes to the self energy $\Sigma(K)$. 
The full $T$-matrix, $t(Q)$, comprises contributions from both condensed pairs, $t_\text{sc}(Q)$, 
and noncondensed pairs, $t_\text{pg}(Q)$, representing the superconducting and pseudogap components, respectively. The former vanishes at $T\ge T_c$, which we shall ignore here. 
The latter is given by $t_\text{pg}(Q)=U/[1+U \chi(Q)]$ under the ladder approximation. Here 
 $\chi(Q)=\frac{1}{2}\sum_K [G_{0\uparrow}(Q-K)G_\downarrow(K)+G_{0\downarrow}(Q-K)G_\uparrow(K)]$ is the  spin-symmetrized pair susceptibility \cite{chien2006PRL}.
Note that the mix of $G_0$ and $G$ in $\chi(Q)$ is a natural result of
the equations of motion approach, where $G_0$ comes from the inverse of the operator $\hat{G}_0^{-1}$ \cite{Kadanoff1961,ChenPhD}.

At $T_\text{c}$, the $T$-matrix contributes to the self-energy as
$\Sigma_\sigma(K)=\Sigma_{\text{pg},\sigma}(K) = \sum_{Q \neq 0}
t_\text{pg}(Q) G_{0\bar{\sigma}}(Q - K)$. Here $t_\text{pg}(Q)$ diverges at $Q = 0$, so that
$\Sigma_\sigma(K)$ is dominated by contributions of $t_\text{pg}(Q)$
in the vicinity of $Q = 0$, and thus can be approximated as
$\Sigma_{\sigma}(K) \approx -\Delta_\text{pg}^2G_{0\bar{\sigma}}(-K)$.
This defines the pseudogap via $\Delta_\text{pg}^2 = -\sum_{Q \neq 0}
t_\text{pg}(Q)$,
with $\Delta\approx \Delta_\text{pg}$ at $T_\text{c}$. With this
approximation, $G(K)$ takes a BCS-like form,
\begin{equation}
G_\sigma(K) = \frac{u_{\mathbf{k}}^2}{\mathrm{i} \omega_n - E_{\mathbf{k}\sigma}} + \frac{v_{\mathbf{k}}^2}{\mathrm{i} \omega_n + E_{\mathbf{k}\bar{\sigma}}}\,.
\end{equation}

The transition $T_\text{c}$ is determined by the Thouless criterion for
pairing instability, $t_\text{pg}^{-1}(0) = 0$, which yields
\begin{equation}
  0=\frac{1}{U}+\sum_{\mathbf{k}}\frac{1-2\bar{f}(E_{\mathbf{k}})}{2E_{\mathbf{k}}}\,,
  \label{eq:gap}
\end{equation}
where $\bar{f}(x)=[f(x+h)+f(x-h)]/2$ and $f(x)=1/(\text{e}^{x/T}+1)$ is the Fermi distribution function. 
Under the fermion number density constraint $n_\sigma = \sum_{K} G_{\sigma}(K)$, 
we obtain
\begin{eqnarray}
n&=&\sum_{\mathbf{k}}\Big[\Big(1-\frac{\xi_{\mathbf{k}}}{E_{\mathbf{k}}}\Big)
 +2\bar{f}(E_{\mathbf{k}})\frac{\xi_{\mathbf{k}}}{E_{\mathbf{k}}}\Big]\,, 
\label{eq:neqa}\\
pn&=&\sum_{\mathbf{k}}\Big[f(E_{\mathbf{k}\uparrow})-f(E_{\mathbf{k}\downarrow})\Big]\,,
\label{eq:neqb}
\end{eqnarray}
with total fermion number density $n = n_\uparrow + n_\downarrow$ and imbalance $p = (n_\uparrow - n_\downarrow) / n$. (As a convention, we refer to the majority fermions as spin up.)

To evaluate the pseudogap via its definition, we need to extract the pair dispersion via the pair propagator or equivalently the $T$-matrix.
After analytical continuation ($\mathrm{i}\Omega_n \rightarrow \Omega+\mathrm{i}0^+$), 
the inverse $T$-matrix $t_\text{pg}^{-1}(Q)$ is Taylor-expanded on the real frequency axis as 
$t_\text{pg}^{-1}(\Omega,\mathbf{q})\approx a_\text{1}\Omega^{2}+a_\text{0}(\Omega-\Omega_{\mathbf{q}}+\mu_\text{p})$, 
where $\Omega_{\mathbf{q}}=2t_\text{B}(3-\cos q_x - \cos q_y- \cos q_z)$, 
 $t_\text{B}$ represents the effective pair hopping matrix element, and $\mu_\text{p}=0$ at $T\le T_c$. 
The pair dispersion is given by 
$\tilde{\Omega}_{\mathbf{q}}=[\sqrt{a_\text{0}^{2}+4a_\text{1}a_\text{0}(\Omega_{\mathbf{q}}-\mu_\text{p})}-a_\text{0}]/2a_\text{1}$, 
which reduces to $\tilde{\Omega}_{\mathbf{q}}=\Omega_{\mathbf{q}}-\mu_\text{p}$ when the $a_1$ term is negligible. 
The coefficients $a_0$, $a_1$, and $t_\text{B}$ are obtained via the Taylor expansion (See  Ref.~\cite{Chen2007} for their counterpart expressions for the 3D continuum case). Using
this inverse $T$-matrix expansion, we arrive at the pseudogap equation
\begin{equation}
  |a_0|\Delta_\text{pg}^{2}=\sum_{\mathbf{q}}\frac{b(\tilde{\Omega}_{\mathbf{q}})}{\sqrt{1+4\dfrac{a_{1}}{a_{0}}\Omega_{\mathbf{q}}}}\,,
  \label{eq:PG}
\end{equation}
where $b(x)=1/(\text{e}^{x/T}-1) = \frac{1}{2} \left(\coth\frac{x}{2T} -1\right)$ is the Bose distribution function. 

Equations (\ref{eq:gap})-(\ref{eq:PG}) form a closed set of self-consistent equations, 
which can be used to solve for $(\mu_\uparrow, \mu_\downarrow, \Delta_\text{pg}, T_\text{c})$ (with $\Delta_\text{sc}=0$).

\subsection*{Comparison between $G_0G$ and $GG$ schemes of the $T$-matrix approximations}

In this section, we explain why we do not choose the $GG$ scheme of the $T$-matrix approximation, despite that it is often referred to as the conserving approximation or self-consistent $T$-matrix approximation. Note that a comparison between $G_0G_0$, $G_0G$ and $GG$ was already given in Ref.~\cite{Levin2010} as well as Ref.~\cite{Morawetz2011}. Here we provide extra reasons.

First, the principle to follow. We have chosen to obey the Heisenberg equations of motion (EoM) as closely as possible rather than simplicity of formality. Our $G_0G$ approach, as first derived by Kadanoff and Martin early in 1961 \cite{Kadanoff1961}, and later extended by Patton \cite{Patton1971}, Levin and coworkers \cite{Maly1999a,ChenPhD}, followed closely the EoM governed by the given Hamiltonian. The $G_0$ in the $T$-matrix comes from the inverse of the operator $\hat{G}_0^{-1}$ in the EoM.
To put simply, the $G_0G$ choice was not constructed diagrammatically by dressing one of the bare fermion Green's functions with the self-energy, but rather a diagrammatic representation of the derived equations. This should be contrasted with the perturbative $G_0G_0$ and the fully dressed $GG$ approximations.

Despite its formal simplicity, the $GG$ formulation does \emph{not} obey the equations of motion, which is the single most important criterion.

The merit of $G_0G$ is obvious. It leads naturally to the BCS form of the pairing gap as a function of temperature, consistent with experimental observations in all superconductors in the superfluid phase so far.
It has been argued in Ref.~\cite{Morawetz2011} that the asymmetric form of the Bethe-Salpeter equation (i.e., our $G_0G$ approach) is a better choice than the symmetric form (i.e., $GG$) for pairing and condensation `` to avoid unphysical repeated collisions is derived as a result of the hierarchical dependencies of correlations.'' It is stated ``Exceeding the parquet approximation we show that an asymmetry appears in the self-consistent propagators.''

\begin{figure}
\centerline{  \includegraphics[width=3.2in]{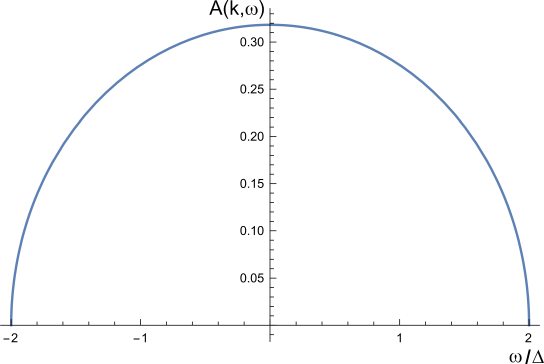}}
  \caption{Spectral function at the Fermi level below $T_c$ in the $GG$ scheme of the $T$-matrix approximation, as per Eq.~(13) of Ref.~\cite{Tchernyshyov1997}. It is not a regular spectral peak.}
  \label{fig:A_GG}
\end{figure}

In contrast, as revealed by Tchernyshyov in 1997~\cite{Tchernyshyov1997}, the $GG$ approach \emph{does not admit a  gap} at the Fermi level in the superfluid phase. One can simply verify this by plotting the spectral function given by Eq.~(13) in Ref.~\cite{Tchernyshyov1997}, which was derived via a pseudogap-like approximation. As shown in Fig.~\ref{fig:A_GG}, the spectral function is a band of continuous distribution instead of a regular spectral peak at the Fermi level.  This absence of an excitation gap at the Fermi level is also confirmed by the rf spectroscopic calculations by Zwerger et al.\ \cite{zwerger2011}. The same $GG$ scheme has been adopted in several recent studies~\cite{Tamaki2008,Frank2018,Kawamura2022}, yet the absence of a Fermi-level gap remains a fundamental limitation. For this reason, we choose not to pursue the $GG$ approach of the $T$-matrix approximation.

\clearpage

